\documentclass[12pt]{article}

\title{Contextualist viewpoint to 
Greenberger-Horne-Zeilinger paradox}

\author{Andrei Khrennikov\\
Department of Mathematics, Statistics and Computer Sciences\\
University of V\"axj\"o, S-35195, Sweden}
\begin{document}
\maketitle

\begin{abstract} We present probabilistic analysis of the Greenberger-Horne-Zeilinger
(GHZ) scheme in the contextualist framework, namely under the assumption that 
distributions of hidden variables depend on settings of measurement devices. On one hand,
we found classes of probability distributions of hidden variables for that the GHZ scheme
does not imply a contradiction between the local realism and quantum formalism. On the other hand,
we found classes of probability distributions of hidden variables for that the GHZ scheme
still induce such a contradiction (despite variations of distributions). It is also demonstrated
that (well known in probability theory) singularity/absolute continuity dichotomy
for probability distributions is closely related to
the GHZ paradox. Our conjecture is that this GHZ-coupling between singularity/absolute continuity dichotomy
and incompatible/compatible measurements might be a general feature of quantum theory.
\end{abstract}

\section{Introduction}

Violations of Bell's inequality [1] by
quantum correlations may be interpreted as an evidence of the impossibility
to use the {\it local realism} in quantum theory (see, for example, [2], [3]).
Despite the general attitude to connect violations of
Bell's inequality with such problems as {\it determinism and locality},
there exists sufficiently strong opposition [4]-[7] to 
such a conclusion. 
We call such an opposition the {\it probability
opposition.} The general viewpoint of adherents of the
probabilistic interpretation of violations of Bell's inequality is
that the derivation of this inequality is based on some
(implicit) probabilistic assumptions.
It seems that theoretical as well as experimental
investigations of the EPR paradox (in particular, Bell's
inequality) must be at least partly reoriented to the
investigation of probabilistic roots of this paradox. 

A new strong argument in the favour of
nonlocal (or nonreal) interpretation of violations of Bell's inequality  was
given by so called Greenberger-Horne-Zeilinger (GHZ) paradox, [8].
The GHZ scheme is based on the probability ${\bf P}=1$ arguments. From the
first point of view all probabilistic circumstances of the GHZ
scheme are so straightforward that there is no more place for probabilistic
counter arguments. There is  rather general opinion that probability does not
play any role in the GHZ scheme: {\it probability ${\bf P}=1$  statements are typically considered
as deterministic statements.} However, the careful probabilistic analysis 
demonstrates that the GHZ paradox has even deeper connection to
foundations of probability theory than Bell's inequality.

In the present paper we study probabilistic structures induced by the combinations
of quantum systems and measurement devices. We consider `integral hidden variables'
$w=(\lambda, \lambda^a, \lambda^b, \lambda^c),$ where $\lambda$ corresponds to a physical 
system (triple of photons in the GHZ framework) and $\lambda^a, \lambda^b, \lambda^c$ correspond to
internal states of measurement devices. On the space of such hidden variables we introduce
probability distribution ${\bf P}_{\phi_1\phi_2\phi_3}$ corresponding to settings 
$\phi_1,\phi_2,\phi_3$ (phase shifts) of measurement apparatuses $A,B,C.$ In fact,
the GHZ used the implicit probabilistic assumption:

${\cal I}${\it Probability distribution ${\bf P}_{\phi_1\phi_2\phi_3}$ does not depend on 
phase shifts $\phi_1,\phi_2,\phi_3: \; \; {\bf P}_{\phi_1\phi_2\phi_3}\equiv {\bf P} $}

It is demonstrated that if ${\cal I}$ does not hold true, then the GHZ considerations do 
not imply a paradox. In fact, our study of the GHZ paradox is nothing than an application
of so called contextualist approach to quantum mechanics. This approach was developed [6].
\footnote{Contextualist approach is characterized by diversity of models, see [6]. 
We do not try to compare or discuss
these models. In fact, my contextualist views were formed on the basis of reading of
papers of W. De Muynck and S. Gudder.} However, we do not only apply the contextualist approach to the GHZ paradox.
We did essentially more. On one hand, we found classes of probability distributions 
of hidden variables in that the GHZ scheme does not imply a contradiction between
the local realism and quantum formalism.
On the other hand, we found classes of probability distributions 
of hidden variables in that the GHZ scheme still implies such a contradiction despite
variations of probability distributions.

Notions of mutually singular and absolutely continuous distributions [9] play an important role
in this framework. It is proved that if the transformation
${\bf P}_{\phi_1\phi_2\phi_3} \to {\bf P}_{\phi_1^\prime\phi_2^\prime\phi_3^\prime}$
is singular, then the GHZ scheme is destroyed; if this transformation is absolutely continuous,
then the GHZ scheme still works. Therefore if (in the contextualist framework) we suppose 
that distributions of hidden variables in measurements of incompatible observables
are absolutely continuous, then via the GHZ considerations we must again obtain
a contradiction between the local realism and quantum formalism. To save the local realism,
we have to assume that incompatible measurements produce probability distributions of hidden
variables which are not absolutely continuous. 

This  coupling (via the GHZ framework) of singularity/absolute continuity  dichotomy (see any
advanced 
textbook in probability theory, for example, [10]) with
incompatible/compatible measurements complementarity induce a conjecture:

{\bf General splitting of physical measurements into quantum/classical measurements
is an exhibition of singularity/absolute continuity  dichotomy in probability theory.}

The singularity/absolute continuity  dichotomy is one of fundamental facts of probability theory [10]:
if the number of degrees of freedom  $N\to \infty,$ then (under quite general assumptions)
probability distributions are either singular or absolutely continuous.

In principle, distributions of hidden variables may fluctuate not only due to a change 
in the measurement arrangement, but also due to statistical variations corresponding to
different runs of an experiment (fluctuations in preparation). This is a hypothesis
on ensemble nonreproducibility, see W. De Baere [5], W. De Muynck, W. De Baere, H. Martens [6]
and the author [5]. The ensemble reproducibility is used as an implicit probabilistic
assumption both in Bell's and GHZ's frameworks: 

${\cal B}$ {\it The probability distribution ${\bf P}$ of hidden variables $\lambda$
(corresponding to physical systems)
in an ensemble $\Omega$ prepared for some measurement is uniquely determined by fixing of a
quantum state $\psi.$ }

So formally, if $\Omega = \Omega_\psi,$ then we can write ${\bf P}_\psi$ instead
of ${\bf P}_{\Omega_\psi}.$ 

We demonstrate that if ${\cal B}$ is violated the GHZ considerations do not imply 
a contradiction between the local realism and quantum formalism.

\section{Probabilistic analysis of the GHZ scheme}

{\bf 1. GHZ arguments.} We briefly repeat GHZ arguments. We shall use an advanced
variant of these arguments, see A. Shimony in [3]. The latter form of the `GHZ-paradox'  is based on 
rather sofisticated probabilistic considerations. It is really not easy to critisize
these considerations from the viewpoint of hidden variables theory. On the other hand,
the original GHZ arguments [8] can be easily destroyed by using the contextualist approach
(and elementary probabilistic reasons), see Remark 1.2 at the end of this section.

There are three different phase shifts, $\varphi_1, \varphi_2,\varphi_3.$ For each fixed setting of 
three shifts, there are three measurements (for three photons which are produced by the down-conversion process 
from a single photon). Results of these measurements are denoted $\rm{  A(\varphi_1), B(\varphi_2), C(\varphi_3)=\pm 1.}$
Quantum formalism predicts that:
\begin{equation}
\label{a}
{\bf P}^{(\psi)}(A(\varphi_1)B(\varphi_2)C(\varphi_3)=1|\varphi_1+\varphi_2+\varphi_3=
{\frac{\pi}{2}},\rm{mod}\;2\pi)=1\;,  
\end{equation}
\begin{equation}
\label{b}
{\bf P}^{(\psi)}(A(\varphi_1)B(\varphi_2)C(\varphi_3)=-1|\varphi_1+\varphi_2+\varphi_3={\frac{\pi}{2}},\rm{mod}\;2\pi)=0\;, 
\end{equation}
\begin{equation}
\label{c}
{\bf P}^{(\psi)}(A(\varphi_1)B(\varphi_2)C(\varphi_3)=1|\varphi_1+\varphi_2+\varphi_3={\frac{3\pi}{2}},\rm{mod}\;2\pi)=0´\;,
\end{equation}
\begin{equation}
\label{d}
{\bf P}^{(\psi)}(A(\varphi_1)B(\varphi_2)C(\varphi_3)=-1|\varphi_1+\varphi_2+\varphi_3={\frac{3\pi}{2}},\rm{mod}\;2\pi)=1\;.
\end{equation}

Let us describe the above experiment by using hidden variables.
The GHZ used so called deterministic hidden variables model in that
by fixing a value $\lambda= \lambda_0$ we fix values of (in general incompatible)
physical observables.

It is assumed that there exists a Kolmorogov probability space $(\Omega, {\cal F}, {\bf P})$ where $\Omega$ 
is the configuration space of hidden variables 
$\lambda$ (it is typically denoted by $\Lambda$
in papers on Bell or GHZ considerations); ${\cal F}$ is a $\sigma$-field of subsets of $\Omega, 
{\bf P}$ is the probability distribution of hidden variables. Physical observables 
$\rm{  A(\varphi_1), B(\varphi_2), C(\varphi_3)}$ are represented by random variables 
on $\Omega:\rm{A=A(\varphi_1, \lambda), B=B(\varphi_2, \lambda), C=C(\varphi_3, \lambda).   }$ Thus (\ref{a}), 
(\ref{b}) and (\ref{c}), (\ref{d}) imply that if $\varphi_1+\varphi_2+\varphi_3={\frac{\pi}{2}},$ 
then 
$\rm{A(\varphi_1, \lambda)B(\varphi_2, \lambda)C(\varphi_3,\lambda)}=1$
for $\lambda \in \Omega^+_{\varphi_1\varphi_2\varphi_3}
\in {\cal F}, \;
{\bf P}(\Omega^+_{\varphi_1\varphi_2\varphi_3})=1,$
and if $\varphi_1+\varphi_2+\varphi_3=
{\frac{3\pi}{2}}, $ then 
$\rm{A(\varphi_1, \lambda)B(\varphi_2, \lambda)C(\varphi_3,\lambda)}=-1$ 
for $\lambda \in \Omega^-_{\varphi_1\varphi_2\varphi_3} \in {\cal F},
{\bf P}(\Omega^-_{\varphi_1\varphi_2\varphi_3})=1.$

Thus we have: 
$A({\frac{\pi}{2}},\lambda) B(0,\lambda) C(0,\lambda) = 1,$ where 
$\lambda \in \Omega^+_{\frac{\pi}{2}00};\; ; \;$
$A(0,\lambda)B({\frac{\pi}{2}},
\lambda)C(0,\lambda)=1, $ where $\lambda\in\Omega^+_{0{\frac{\pi}{2}}0},$ and
$A(0,\lambda) B(0,\lambda) C(\frac{\pi}{2},\lambda)=1,$ where 
$\lambda\in\Omega^+_{00\frac{\pi}{2}}.$

Set 
$
\Sigma^+=
\Omega^+_{\frac{\pi}{2}00}\cap\Omega^+_{0\frac{\pi}{2}0}\cap\Omega^+_{00\frac{\pi}{2}}.
$
As ${\bf P}(\Omega^+_{\frac{\pi}{2}00})={\bf P}(\Omega^+_{0\frac{\pi}{2}0})={\bf P}(\Omega^+_{00\frac{\pi}{2}})=1,$ 
we obtain that
${\bf P}(\Sigma^+)=1.$

We have that, for each $\lambda\in \Sigma^+,$
$A (\frac{\pi}{2},\lambda) B (\frac{\pi}{2},\lambda) C(\frac{\pi}{2},\lambda)=1.
$
Thus $\sum^+ \subset \Omega^+_{\frac{\pi}{2}\frac{\pi}{2}\frac{\pi}{2}}= \Omega\setminus \Omega^-_{\frac{\pi}{2}\frac{\pi}{2}\frac{\pi}{2}}.$ 
But ${\bf P}(\Omega^+_{\frac{\pi}{2}\frac{\pi}{2}\frac{\pi}{2}})=
1-{\bf P}(\Omega^-_{\frac{\pi}{2}\frac{\pi}{2}\frac{\pi}{2}})=0$  and,
hence, 
${\bf P}(\Sigma^+)=0.$
This is the GHZ `paradox'.
The standard inference is that we cannot use local hidden variables, because of (\ref{d}).

{\bf 2. Contextualist model.} Here we could not suppose that
by fixing a value $\lambda= \lambda_0$ of the hidden variable of a quantum system
we fix values of (in general incompatible)
physical observables. Internal states of measurement devices must be also 
taken into account (see, for example, J.Bell [11]).

Denote by $\rm{\Lambda, \Lambda_{\varphi_1}^a, \Lambda_{\varphi_2}^b, \Lambda_{\varphi_3}^c  }$ spaces of
hidden variables, respectively, for triples of photons and measurement devices
$\rm{A(\varphi_1), B(\varphi_2), C(\varphi_3)}.$ 

Here
$\rm{A(\varphi_1)=A(\varphi_1, \lambda, \lambda^a), B(\varphi_1)=B(\varphi^1, \lambda, \lambda^b),
 C(\varphi_1, \lambda, \lambda^c)}.$
Thus (\ref{a}), (\ref{b}) and (\ref{c}), (\ref{d}) imply that if $\varphi_1+ \varphi_2+ \varphi_3=\frac{\pi}{2},$ then
\begin{equation}
\label{a1}
\rm{A(\varphi_1, \lambda, \lambda^a) B(\varphi_2, \lambda, \lambda^b) C(\varphi_3, \lambda,
\lambda^c)=1}
\end{equation}
for
$\rm{w=(\lambda, \lambda^a, \lambda^b, \lambda^c)\in\Omega^+_{\varphi_1\varphi_2\varphi_3}\in
{\cal F}_{\varphi_1\varphi_2\varphi_3}}$ and
\begin{equation}
\label{a2}
{\bf P}_{\varphi_1\varphi_2\varphi_3}(\Omega^+_{\varphi_1\varphi_2\varphi_3})=1,
\end{equation}
 and if
$\varphi_1+ \varphi_2+ \varphi_3=\frac{3^\pi}{2},$ then
\begin{equation}
\label{a3}
\rm{A(\varphi_1, \lambda, \lambda^a) B(\varphi_2, \lambda, \lambda^b) C(\varphi_3, \lambda,
\lambda^c)= - 1}
\end{equation}
for
$\rm{w=(\lambda, \lambda^a, \lambda^b
\lambda^c)\in\Omega^-_{\varphi_1\varphi_2\varphi_3}\in {\cal F}_{\varphi_1\varphi_2\varphi_3}}$ and
\begin{equation}
\label{a4}
{\bf P}_{\varphi_1\varphi_2\varphi_3}(\Omega^-_{\varphi_1\varphi_2\varphi_3})=1 .
\end{equation}

The total space of
hidden variables for the system of quantum particles and measurement apparatuses is the set
$\rm{\Omega_{\varphi_1\varphi_2\varphi_3}=\Lambda \times \Lambda^a_{\varphi_1} \times
 \Lambda^b_{\varphi_2}
\times \Lambda^c_{\varphi_3}}.$ We denote by the symbol $\rm{{\cal F}_{\varphi_1\varphi_2\varphi_3}}$ a
$\sigma$-field of subsets of $\Omega_{\varphi_1\varphi_2\varphi_3}.$ The
${\bf P}_{\varphi_1\varphi_2\varphi_3}$
is the probability distribution of hidden variables $\rm{w=(\lambda, \lambda^a, \lambda^b,
\lambda^c).}$

{\bf{Remark 1.1.}} It is natural that the distribution of hidden variables w depends on the configuration
$(\varphi_1\varphi_2\varphi_3)$ of phase shifts. In fact, we should use the symbol $\rm{w_{\varphi_
1\varphi_2\varphi_3}=(\lambda,\lambda^a_{\varphi_1},\lambda^b_{\varphi_2},\lambda^c_{\varphi_3})}
$ to denote this hidden multivariable. 

Thus we have
\begin{equation}
\label{a5}
\rm{A(\frac{\pi}{2},\lambda, \lambda^a)B(0,\lambda,\lambda^b)C(0, \lambda, \lambda^c)=1, w\in \Omega^+_{
\frac{\pi}{2}00};}
\end{equation}
\begin{equation}
\label{a6}
\rm{A(0,\lambda, \lambda^a)B(\frac{\pi}{2},\lambda,\lambda^b)C(0, \lambda, \lambda^c)=1, w\in \Omega^+_{0
\frac{\pi}{2}0};}
\end{equation}
\begin{equation}
\label{a7}
\rm{A(0,\lambda, \lambda^a)B(0,\lambda,\lambda^b)C(\frac{\pi}{2}, \lambda, \lambda^c)=1, w\in \Omega^+_{
00\frac{\pi}{2}};}
\end{equation}
\begin{equation}
\label{a8}
\rm{A(\frac{\pi}{2},\lambda, \lambda^a)B(\frac{\pi}{2},\lambda,\lambda^b)C(\frac{\pi}{2}, \lambda,
\lambda^c)=-1, w\in \Omega^-_{\frac{\pi}{2}\frac{\pi}{2}\frac{\pi}{2}};}
\end{equation}
\begin{equation}
\label{a9}
\rm{{\bf P}_{\frac{\pi}{2}00}(\Omega^+_{\frac{\pi}{2}00})=1,{\bf P}_{0\frac{\pi}{2}0}(\Omega^+_{0\frac{
\pi}{2}0})=1,{\bf P}_{00\frac{\pi}{2}}(\Omega^+_{00\frac{\pi}{2}})=1; }
\end{equation}
\begin{equation}
\label{a10}
\rm{{\bf
P}_{\frac{\pi}{2}\frac{\pi}{2}\frac{\pi}{2}}(\Omega^-_{\frac{\pi}{2}\frac{\pi}{2}\frac{\pi}{2}}})=1 \;.
\end{equation}

The following two assumptions will play an important role in our further considerations:

(A) The space of hidden variables $\Omega_{\varphi_1\varphi_2\varphi_3}$ does not depend on shifts
$\varphi_1,\varphi_2,\varphi_3$ (the sets of possible microstates of apparatuses do not depend on
shifts).

(B) The distribution of hidden variables ${\bf P}_{\varphi_1,\varphi_2,\varphi_3}$ does not depend on
shifts $\varphi_1,\varphi_2,\varphi_3$

Under assumption (A) we can set $\Omega=\Omega_{\varphi_1\varphi_2\varphi_3}.$ Here we can define the set
$\Sigma^+=\Omega^+_{\frac{\pi}{2}00}\cap\Omega^+_{0\frac{\pi}{2}0}\cap\Omega^+_{00\frac{\pi}{2}}.$

It is evident that
$\rm{\Sigma^+ \subset \Omega^+_{\frac{\pi}{2}\frac{\pi}{2}\frac{\pi}{2}}}.
$
Thus (\ref{a10}) implies that
\begin{equation}
\label{a13}
{\bf P}_{\frac{\pi}{2}\frac{\pi}{2}\frac{\pi}{2}}(\Sigma^+)=0.
\end{equation}

Under assumption (A)+(B) we can set $\Omega=\Omega_{\varphi_1\varphi_2\varphi_3}$ and
${\bf P}={\bf P}_{\varphi_1\varphi_2\varphi_3}.$ We can omit indexes of probability distributions in (
\ref{a9}) and (\ref{a13}) and obtain ${\bf P}(\sum^+)=1$ and
${\bf P}(\sum^+)=0.$ This is the GHZ paradox.

To obtain the GHZ paradox, we must assume (A) and (B). The assumption (A) seems to be quite
natural: even if some hidden parameters w for shifts configuration $\varphi_1,\varphi_2,\varphi_3$ are
eliminated by other shifts configuration $\varphi_1^\prime,\varphi_2^\prime,\varphi_3^\prime,$ we can
still assume that they belong to the space
$\Omega_{\varphi_1^\prime,\varphi_2^\prime,\varphi_3^\prime,}$ by setting
${\bf P}_{\varphi_1^\prime,\varphi_2^\prime,\varphi_3^\prime,}(\rm{w})=0.$ However, we have to recognize
that the assumption (B) has no physical justification at the present level of quantum experiments:

1). It
seems that by changing the experimental arrangement (configuration of phase shifts) we change the
distribution of hidden variables (corresponding to quantum particles+measurement devices), so
${\bf P}_{\varphi_1,\varphi_2,\varphi_3}\not={\bf
 P}_{\varphi_1^\prime,\varphi_2^\prime,\varphi_3^\prime}.$

 2). Distributions used in GHZ considerations are induced by four different runs of the experiment. It
 may be that distributions of hidden variables (even for $\lambda$) fluctuate from run to run. This is
 the hypothesis of nonreproducibility (see [5]). At the moment we have neither arguments against
 this hypothesis nor in favour of this hypothesis. 
 
 Further considerations will be performed under
 assumption (A).

 {\bf Remark 1.2.} (The original GHZ arguments). The original GHZ scheme [8] was based on the consideration
 of angles, $\alpha, \beta, \gamma, \delta.$ We follow this scheme. We set 
 $\Pi(\alpha, \beta, \gamma, \delta) =
 A(\alpha, \lambda) B(\beta, \lambda) C(\gamma,  \lambda) D(\delta,  \lambda),$
 where $A,B,C,D$ are physical observables considered in [8].
 Greenberger, Horne and Zeilinger obtained the following conditions:
 \begin{equation}
  \label{T1}
  \Pi(\alpha, \beta, \gamma, \delta) =1, \; 
  \alpha + \beta + \gamma +  \delta= 0,
 \end{equation}
 and 
 \begin{equation}
  \label{T2}
  \Pi(\alpha, \beta, \gamma, \delta) = - 1, \; 
  \alpha + \beta + \gamma +  \delta= \pi.
 \end{equation}
 Then they remarked: {\small " But it turns out that there is no way to satisfy this condition. It is too
 restrictive,
 because we can continuously vary two of the parameters while keeping the other two constant. 
 This leads to the conclusion that $A=B=C=D= constant.$ But it is impossible, since the product
 sometimes equals  +1 and sometimes equals -1. This is true for any value of $\lambda$ so that there is 
 no need to integrate over it,"} [8], p.72.
 
 Unfortunetely these considerations are based on a rather elementary misunderstanding of the notion of 
 probability 1, see,  for example, [9], [10]. In fact, the probability 1 arguments need not imply that 
 something  "{\bf is true for any value of $\lambda$}." Such argumnets only imply that, for example, 
 (\ref{T1}) holds true for $\lambda$ belonging to a set $\Omega^+_{\alpha\beta\gamma\delta}$
 which has the probability measure 1. $\Omega^+_{\alpha\beta\gamma\delta}$ may depend on the parameters
 $\alpha, \beta, \gamma, \delta.$ Moreover, this is the typical situation even in `classical probabilistic
 models', see, for example, [10]. Therefore we could not vary the parameters $\alpha, \beta, \gamma, \delta$
 for a fixed value of $\lambda.$ Thus there are no reasons to suppose (as it was done by 
 Greenberger, Horne and Zeilinger) that $A=B=C=D= constant.$ Finally, we remark that 
 the assumption that the set $\Omega^+_{\alpha\beta\gamma\delta}$ depends on the experimental
 settings $\alpha, \beta, \gamma, \delta$ is nothing than a  contextualist assumption.

 \section{GHZ scheme for absolutely continuous and singular variations of probability distributions of hidden
 variables}

{\bf{1. Absolutely continuous and singular probability distributions.}} 
Let ${\bf P}^\prime$ and ${\bf P}^{\prime\prime}$ be two probability measures. $
{\bf P}^{\prime\prime}$ is absolutely continuous with respect to ${\bf P}^\prime$ if ${\bf P}^{\prime\prime}(E)=0$
whenever ${\bf P}^\prime (E)=0, E \in {\cal F} \; ({\bf P}^{\prime\prime}<<{\bf P}^\prime).  
{\bf P}^{\prime\prime}$ is singular with respect to ${\bf P}^\prime$ if there is a set $E \in {\cal F}$ 
such that ${\bf P}^{\prime\prime}(E)=1$ and ${\bf P}^\prime (E)=0 \;({\bf P}^{\prime\prime}\perp{\bf P}^\prime$).
If ${\bf P}^{\prime\prime}$ and ${\bf P}^{\prime}$ are mutually absolutely continuous, they are called
equivalent.

{\bf 2. GHZ paradox for equivalent probability distributions of hidden variables.}
Suppose that different settings $\varphi_1,\varphi_2,\varphi_3$ 
 (and different runs of the experiment)
 produce in general different probability distributions 
 ${\bf P}_{\varphi_1,\varphi_2,\varphi_3},$ but
 they are absolutely continuous with respect to each other:
 ${\bf P}_{\varphi_1,\varphi_2,\varphi_3}$ is equivalent to ${\bf
  P}_{\varphi_1^\prime,\varphi_2^\prime,\varphi_3^\prime}.$ Thus
 \begin{equation}
 \label{b1}
 {\bf P}_{\varphi_1^\prime,\varphi_2^\prime,\varphi_3^\prime}(dw)=
 f(w;\varphi_1^\prime\varphi_2^\prime\varphi_3^\prime/\varphi_1\varphi_2\varphi_3){\bf P}_{\varphi_1\varphi_2\varphi_3}(dw),
 \end{equation}
  where $f$ is the density function.

  {\bf `Theorem'.} {\it Suppose that probability distributions of hidden variables
 ${\bf P}_{\varphi_1^\prime,\varphi_2^\prime,\varphi_3^\prime}$ and ${\bf
  P}_{\varphi_1,\varphi_2,\varphi_3}$ are equivalent for all possible settings 
  of measurement devices in the GHZ scheme.
 Then
 GHZ arguments imply that quantum formalism and local realism are incompatible.}

 {\bf Proof.} As
 ${\bf P}_{\frac{\pi}{2}\frac{\pi}{2}\frac{\pi}{2}}$ is absolutely continuous with respect to
 ${\bf P}_{\frac{\pi}{2}00}, {\bf P}_{0\frac{\pi}{2}0}$ and ${\bf P}_{00\frac{\pi}{2}},$
 we obtain that
 ${\bf P}_{\frac{\pi}{2}00}(\Omega^+_{\frac{\pi}{2}00})=1\rightarrow{\bf
  P}_{\frac{\pi}{2}\frac{\pi}{2}\frac{\pi}{2}}(\Omega^+_{\frac{\pi}{2}00})=1,\ldots,{\bf
  P}_{00\frac{\pi}{2}}(\Omega^+_{00\frac{\pi}{2}})=1\rightarrow{\bf
  P}_{\frac{\pi}{2}\frac{\pi}{2}\frac{\pi}{2}}(\Omega^+_{00\frac{\pi}{2}})=1.$
 Thus ${\bf P}_{\frac{\pi}{2}\frac{\pi}{2}\frac{\pi}{2}}(\sum^+)=1.$ On the other hand, as usual, we have
 ${\bf P}_{\frac{\pi}{2}\frac{\pi}{2}\frac{\pi}{2}}(\sum^+)=0.$

{\bf {3. No GHZ inference for singular probability distributions of hidden variables.}} 
Suppose that ensemble fluctuations produce singular distributions of hidden 
variables. Thus ${\bf P}_{\varphi_1,\varphi_2,\varphi_3}\perp
{\bf P}_{\varphi_1^\prime,\varphi_2^\prime,\varphi_3^\prime}.$

Suppose that $\Omega^+_{\frac{\pi}{2}00}, \Omega^+_{0\frac{\pi}{2}0}, \Omega^+_{00\frac{\pi}{2}}$ play the role of 
the set $E$ in the definition of singularity for 
distributions ${\bf P}_{\frac{\pi}{2}00}$ and ${\bf P}_{\frac{\pi}{2}\frac{\pi}{2}\frac{\pi}{2}},$ 
${\bf P}_{0\frac{\pi}{2}0}$ and ${\bf P}_{\frac{\pi}{2}\frac{\pi}{2}\frac{\pi}{2}},$ ${\bf P}_{00\frac{\pi}{2}}$ and 
${\bf P}_{\frac{\pi}{2}\frac{\pi}{2}\frac{\pi}{2}},$ respectively.
Then we have:

${\bf P}_{\frac{\pi}{2}00}(\Omega^+_{\frac{\pi}{2}00})=1$ and ${\bf P}_{\frac{\pi}{2}\frac{\pi}{2}\frac{\pi}{2}}(\Omega^+_{\frac{\pi}{2}00})=0;$

${\bf P}_{0\frac{\pi}{2}0}(\Omega^+_{0\frac{\pi}{2}0})=1$ and ${\bf P}_{\frac{\pi}{2}\frac{\pi}{2}\frac{\pi}{2}}(\Omega^+_{0\frac{\pi}{2}0})=0;$

${\bf P}_{00\frac{\pi}{2}}(\Omega^+_{00\frac{\pi}{2}})=1$ and ${\bf P}_{\frac{\pi}{2}\frac{\pi}{2}\frac{\pi}{2}}(\Omega^+_{00\frac{\pi}{2}})=0.$

Therefore ${\bf P}_{\frac{\pi}{2}\frac{\pi}{2}\frac{\pi}{2}}(\Sigma^+)=0.$ Thus there is no GHZ `paradox'.

{\bf{4. Infinite-dimensional spaces of hidden variables and GHZ paradox.}} Let $\Omega$ be an infinite 
dimensional linear space. Singularity of probability measures on $\Omega$ is quite typical. We shall 
consider the example of Gaussian distributions on a Hilbert space $\Omega$ of hidden variables (here ${\cal F}$
is a $\sigma$-field of Borel sets). We shall demonstrate that singularity of 
Gaussian probabilities can be induced by negligibly small perturbations of parameters of these 
distributions. Thus, in principle, we may obtain singular distributions of hidden variables in 
different runs of the experiment due to negligibly small fluctuations of parameters in the 
preparation device (as well as measurement devices). Let $\xi(\omega)$ and 
$\xi^\prime( \omega)$ be Gaussian random variables in the Hilbert space $\Omega$ with 
mean values $a$ and $a^\prime (\in \Omega)$ 
and covariation operators B and B$^\prime$, respectively (see, for example, [12]).

First we consider the case in that ensemble fluctuations can change only mean 
values: $a^\prime = a+ \delta a,$ where $\delta a \in \Omega$ 
is a perturbation of $a \in \Omega.$ So $B = B^\prime.$ 
Let $\{e_j \}_{j=1}^\infty$ be  
the orthonormal basis in $\Omega$ consisting of eigenvectors of the covariation operator 
$B:\; B e_j= b_j e_j, 
j=1, \ldots, \infty.$ We suppose that $B > 0,$ so $b_j>0.$ 
Let ${\bf P} ={\bf P}_\xi$ and ${\bf P}^\prime = {\bf P}_{\xi^\prime}$ 
be probability distributions of Gaussian random variable $\xi$ and $\xi^\prime.$ They are singular if
\begin{equation}
\label{q1}
\sum_{j=1}^\infty \frac{(\delta a_j)^2}{b_j} = \infty, \; \;
\delta a=(\delta a_1, \ldots,  \delta a_N, \ldots)\;.
\end{equation}
For example, suppose that $\delta a_j = \epsilon
\sqrt{\frac{b_j}{j}}$, where $\epsilon>0$ is an arbitrary small constant. 
Then ${\bf P}\perp{\bf P}^\prime.$ We remark that the covariation operator $B$ 
of a Gaussian measure is a nuclear operator in the Hilbert space $\Omega.$ 
Thus $\sum_{j=1}^\infty b_j< \infty.$ 
So $\delta a_j \rightarrow 0, j\rightarrow\infty.$ For example, 
let $\rm{b_j=\frac{1}{j^2}}$ and $\epsilon=10^{-100}.$ 
The perturbation $\delta a_j =10^{-100}/j^{3/2}, j=1,2,\ldots,$ would imply singularity. 
Therefore, to escape singularity in different runs, we must have extremely good statistical 
reproducibility. 

We recall that $b_j= E (\xi_j - a_j)^2
=\int (e_j, \lambda - a)^2 {\bf P}(d\lambda)$ is dispersion of  the (Gaussian) 
random variable $\xi_j=(e_j,\xi)$ (here $a_j = (e_j,a)).$
Relation (\ref{q1}) 
implies that if we increase the sharpness of the distribution of hidden variables 
$\rm{b_j\rightarrow\gamma_j b_j, j=1, \ldots,\gamma_j<<1,}$ then we must decrease the ranges 
of perturbations $\delta a_j$ to escape singularity. Hence, if we approach the domain 
of eigenstates for hidden variables $\lambda:\; b_j\approx 0,$ then we have to have 100\% reproducibility 
of statistical distributions of hidden variables ($\delta a_j\approx 0$) 
to escape the singularity. 

{\bf{Conclusion.}} {\it{A sharp preparation of hidden variables practically definitely implies 
the singularity of probability distributions of hidden variables for different runs of an experiment.}}

We now consider the effect of fluctuations of the parameter $B: B^\prime=B+ \delta B$, where 
$\delta B$ is a perturbation of the covariance operator $\delta B.$ 
We study the simplest case in that the operator $\delta B$  is diagonal 
in the basis $\{e_j \}_{j=1}^\infty$ (consisting of eigenvectors of $B).$ 
We exclude from considerations the case $[B, B^\prime]\not=0$ (which may be interesting). 
So let $\delta B e_j = \delta b_j e_j, j=1,\ldots,\infty, \delta b_j \geq 0.$
It is assumed that $a^\prime = a.$ 
Gaussian distributions ${\bf P}$ and ${\bf P}^\prime$ are singular if (see [12]):
\begin{equation}
\label{q2}
\sum_{j=1}^\infty \frac{(\delta b_j)^2}{b_j^2}= \infty \;.
\end{equation}

For example, suppose that $\delta b_j
=\frac{\epsilon b_j}{\sqrt{j}}:$ where $\epsilon>0$ is an arbitrary small constant. 
Then ${\bf P}\perp{\bf P}^\prime.$ Thus singularity of distributions corresponding for different 
runs can be induced by negligibly small fluctuations of dispersion parameters. We again observe 
that if $\xi(\omega)$ gives a sharp distribution of hidden variables, namely 
$b_j \approx 0$ for all $j,$ then, to escape singularity, we need to have practically precise reproducibility:
$\delta b_j\approx 0$ for all $j.$

{\bf{5. Physical meaning of infinite dimensional spaces of hidden variables.}} 
There are a few possible
 sources of the infinite dimension of the space of hidden variables $\Omega:$ (1) Extremely complex
 structure (from the micro viewpoint) of measurement apparatuses. (2) It may be that physical observables
 have to be described as functions of the whole trajectories of hidden variables (for quantum systems and
 measurement apparatuses) in the process of interaction:
 $\rm{A=A(\lambda(\cdot),\lambda^a(\cdot)),B=B(\lambda(\cdot), \lambda^b(\cdot)), 
 C=C(\lambda(\cdot), \lambda^c(\cdot)).}$ These
 trajectories are nothing than infinite dimensional hidden variables
 (compare with De Muynck and Stekelenborg in [6]). (3)(Bohm-Hiley conjecture, [13])
 Quantum particles might have an extremely complex internal structure. Such a complexity can be described
 by infinite dimensional spaces of hidden variables.

 In this paper we consider assumption (1). Each measurement apparatus consists of a huge number of
 quantum systems. If
 each quantum system can be described by a hidden variable $\rm{\lambda_j^a\in R},$ then the hidden variable
 of the whole apparatus
 $\rm{\lambda^a=(\lambda_j^a)^N_{j=1}, N\rightarrow \infty.}$ The assumption on the Gaussian distribution
 of hidden variables for an apparatus is quite natural:
 the concrete setting of an apparatus is created by the concentration of parameters to some mean value
 $\rm{\lambda_0^a=(\lambda_{j0}^a)}$ (corresponding to this setting).

 \section{Singularity/equivalence dichotomy, the GHZ paradox, the principle of complementarity.}

 Two Gaussian measures on an infinite dimensional space is either singular or equivalent (Hajek-Feldman
 dichotomy, [12]). Our considerations demonstrated that this mathematical fact has the close relation to
 foundations of quantum mechanics. If we use the GHZ scheme, but do not apply to nonlocality or
 determinism, then it seems that to escape the GHZ paradox we have to assume that
 quantum measurement/preparation procedure generates singular distributions of hidden parameters for
 incompatible measurements; classical measurement/preparation procedure generate only equivalent
 distributions of hidden parameters.

 We can speculate that the principle of complementarity is nothing than the exhibition of singularity of
 probability distributions of hidden variables for incompatible measurement.

 Thus in classical measurements we always obtain equivalent probability distributions, in quantum
 measurements there are settings having singular probability distributions.

 In fact, singularity/equivalence dichotomy is not a property of only Gaussian distributions on infinite-
 dimensional spaces. We have the general Kakutani dichotomy [10]:

{\bf Theorem.} {\it Let $\xi=(\xi_1, \xi_2, \ldots, \xi_{\rm{n}}, \ldots)$ and
 $\eta=(\eta_1, \eta_2, \ldots, \eta_{\rm{n}}, \ldots)$ be sequences of independent random variables
 for which ${\bf P}_{(\eta_1,\ldots,\eta_{\rm{n}}})$
 $<<{\bf P}_{(\xi_1, \ldots, \xi_{\rm{n}})}$ for
 $\rm{n}\geq1.$ Then either ${\bf P}_\eta<<{\bf P}_\xi$ or
 ${\bf P}_\eta\perp{\bf P}_\xi.$}

 {\bf Conclusion.} {\it The rigorous hidden variables
 description of the GHZ measurements demonstrates that there are two classes of preparation/measurement
 procedures: quantum (which may produce singular probability distributions) and classical (which always
 produce absolutely continuous probability distributions).} \footnote{We assume that the preparation/
 measurement procedure depends on a huge (practically infinite) number of hidden parameters.}

Despite the general opinion, the hidden variables description need not imply the reduction
 of `quantum reality' to `classical reality'. Although both realities can be described by deterministic
 hidden variables, there is the crucial difference in behaviour of probability distributions. It seems
 that we have found the origin of this difference: This is singularity/equivalence dichotomy which is a
 general property of the large class of distributions of random variables.

 \section {GHZ scheme in the presence of ensemble fluctuations.}

 In this section we obtain the estimate
 of the measure of ensemble fluctuations $\epsilon$ which is induced by the GHZ scheme. The following
 considerations can be interesting only for models of hidden variables which do not have singularity/
 equivalence dichotomy.

 {\bf 1. Metric on the space of measures.} 
 Let $\mu$ be a signed measure defined on a $\sigma$-algebra ${\cal F}$
 (of subsets of $\Omega).$ Let $\mu=\mu^+ -
 \mu^-, $ where $\mu^+, \mu^-$ are positive measures, be {\it the Jordan decomposition}
 of $\mu,$ see, for example, [9]. The {\it total variation} of $\mu$ is defined as
 $\Vert \mu\Vert= \mu^+ (\Omega) + \mu^- (\Omega).$ 
 Let $\mu$ be a discrete signed measure which is concentrated on a sequence of points
 $\{ \lambda_j \}_{j=1}^\infty,$ namely $\mu(A) = \sum_{\lambda_j \in A} \mu(\lambda_j).$
 Here $\Vert \mu \Vert = \sum_{j=1}^\infty \vert \mu(\lambda_j) \vert .$ Let $\mu$ be
 a signed measure that is absolutely continuous with respect to a positive measure $\nu:\;
 \mu(d\omega)= f(\omega) \nu(d \omega),$ where $f: \Omega \to {\bf R}$ is a $\nu$-integrable 
 function. Here $\Vert \mu \Vert= \int_\Omega \vert f(\omega)\vert \nu(d \omega).$
 
 Denote the space of all (signed) measures on ${\cal F}$ by the symbol ${\cal M}.$ Set
 $\rho(\mu_1, \mu_2) = \Vert \mu_1 - \mu_2 \Vert.$ This is a metric on the space ${\cal M}$
 (and ${\cal M}$ is complete). 
 
 {\bf 2. Probability invariant.} Let $\psi$ be a quantum state. Denote the family of all probability
 distributions of hidden variables corresponding to $\psi$ by the symbol $T_\psi.$ Thus, for different runs
 corresponding to $\psi,$ we prepare in general distinct elements of $T_\psi$
 (if Bell's implicit assumption ${\cal B}$ is true, then $T_\psi$ must be a singleton).
 Set
 \begin{equation}
 \label{E}
 \epsilon_\psi= \sup \{ \rho({\bf P}_1, {\bf P}_2): {\bf P}_1, {\bf P}_2 \in T_\psi\}\; .
 \end{equation}
 This is the probability invariant of the quantum state $\psi.$
 
 {\bf 3. Influences of ensemble fluctuations.} We have

 $0={\bf P}_{\frac{\pi}{2}\frac{\pi}{2}\frac{\pi}{2}}(\Sigma^+)=1-{\bf
  P}_{\frac{\pi}{2}\frac{\pi}{2}\frac{\pi}{2}}(\bar{\Sigma}^+)\geq1-{\bf
  P}_{\frac{\pi}{2}\frac{\pi}{2}\frac{\pi}{2}}(\bar{\Omega}^+_{\frac{\pi}{2}00})-{\bf
  P}_{\frac{\pi}{2}\frac{\pi}{2}\frac{\pi}{2}}(\bar{\Omega}^+_{0\frac{\pi}{2}0})-{\bf
  P}_{\frac{\pi}{2}\frac{\pi}{2}\frac{\pi}{2}}(\bar{\Omega}^+_{00\frac{\pi}{2}})\geq1-{\bf
  P}_{\frac{\pi}{2}00}(\bar{\Omega}^+_{\frac{\pi}{2}00})-{\bf
  P}_{0\frac{\pi}{2}0}(\bar{\Omega}^+_{0\frac{\pi}{2}0})-{\bf
  P}_{00\frac{\pi}{2}}(\bar{\Omega}^+_{00\frac{\pi}{2}})-3\epsilon=1-3 \epsilon.$ 
  For a set $D,$ the symbol $\bar{D}$ denotes the complement of $D.$ So
\begin{equation}
\label{t1}
\epsilon \geq 1/3 .
\end{equation}

 Thus if the measure of ensemble fluctuation $\epsilon$ is larger than 1/3, the GHZ scheme does not imply
 a contradiction between quantum formalism and local realism.

 {\bf Example.} {\small Let
 $\rm{\Omega=\{w_1, \ldots, w_N\}}$ and let ${\bf P}$ and ${\bf P}^\prime$ be two discrete probability
 distributions: $\rm{{\bf P}(w_j)={\bf P}_j}$ and $\rm{{\bf P}^\prime(w_j)={\bf P}_j^\prime}.$ Let
 $\rm{|{\bf P}_j-{\bf P}_j^\prime|= \delta.}$ Then $\rho({\bf P}, {\bf P}^\prime) =N \delta.$ If
 $\delta N \geq 1/3,$ i.e.,
 $\delta \geq \frac{1}{3N},$ then there is no contradiction (via the GHZ scheme) between quantum
 formalism
 and local realism. If $N >>1,$ then the presence of negligibly small ensemble fluctuations destroys
 the GHZ arguments.}

\medskip

{\bf References}

\medskip

[1] J.S. Bell,  Rev. Mod. Phys., {\bf 38}, 447--452 (1966).
J. S. Bell, {\it Speakable and unspeakable in quantum mechanics.}
Cambridge Univ. Press (1987).

[2]  J.F. Clauser , M.A. Horne, A. Shimony, R. A. Holt,
Phys. Rev. Letters, {\bf 49} (1969), 1804-1806´;
J.F. Clauser ,  A. Shimony,  Rep. Progr.Phys.,
{\bf 41} 1881-1901 (1978).
A. Aspect,  J. Dalibard,  G. Roger, 
Phys. Rev. Lett., {\bf 49}, 1804-1807 (1982);
D. Home,  F. Selleri, Nuovo Cim. Rivista, {\bf 14},
2--176 (1991). H. P. Stapp, Phys. Rev., D, {\bf 3}, 1303-1320 (1971);
P.H. Eberhard, Il Nuovo Cimento, B, {\bf 38}, N.1, 75-80(1977); Phys. Rev. Letters,
{\bf 49}, 1474-1477 (1982);
A. Peres,  Am. J. of Physics, {\bf 46}, 745-750 (1978).

[3] B. d'Espagnat, {\it Veiled Reality. An anlysis of present-day
quantum mechanical concepts.} Addison-Wesley(1995).
A. Shimony, {\it Search for a naturalistic world view.} Cambridge Univ. Press (1993).

[4] L. de Broglie, {\it La thermodynamique de la particule isolee.} Gauthier-Villars, Paris,
1964; G. Lochak, Found. Physics, {\bf 6´}, 173-184 (1976);

A. Fine,  Phys. Rev. Letters, {\bf 48}, 291--295 (1982);
P. Rastal, Found. Phys., {\bf 13}, 555 (1983).

[5] W. De Baere,  Lett. Nuovo Cimento, {\bf 39}, 234-238 (1984);
{\bf 25}, 2397- 2401 (1984). 

A. Yu Khrennikov, Il Nuovo Cimento, {\bf 115}(2000),179; 
{\it Interpretations of Probablity} (VSP Int. Sc. Publ.), 1999.
A. Yu. Khrennikov, A perturbation of CHSH inequality induced by fluctuations of ensemble distributions.
To be published in JMP.

[6] L. Accardi, 
Phys. Rep., {\bf 77}, 169-192 (1981).
L. Accardi, A. Fedullo, 
Lettere al Nuovo Cimento {\bf 34} 161-172  (1982).

I. Pitowsky,  Phys. Rev. Lett, {\bf 48}, N.10, 1299-1302 (1982);
Phys. Rev. D, {\bf 27}, N.10, 2316-2326 (1983);
S.P. Gudder,  J. Math Phys., {\bf 25}, 2397- 2401 (1984);

Accardi  L., The probabilistic roots of the quantum mechanical paradoxes.
{\it The wave--particle dualism. A tribute to Louis de Broglie on his 90th 
Birthday}, Edited by S. Diner, D. Fargue, G. Lochak and F. Selleri.
D. Reidel Publ. Company, Dordrecht, 47--55(1984);

W. De Muynck, J.T. Stekelenborg,  Annalen der Physik, {\bf 45},
N.7, 222-234 (1988).

W. De Muynck and W. De Baere W.,
Ann. Israel Phys. Soc., {\bf 12}, 1-22 (1996);
W. De Muynck, W. De Baere, H. Marten,
Found. of Physics, {\bf 24}, 1589--1663 (1994);

[7] A. Yu. Khrennikov,  Dokl. Akad. Nauk SSSR, ser. Matem.,
{\bf 322}, No. 6, 1075--1079 (1992); J. Math. Phys., {\bf 32}, No. 4, 932--937 (1991);
Physics Letters A, {\bf 200}, 119--223 (1995);
Physica A, {\bf 215}, 577--587 (1995);   Int. J. Theor. Phys., {\bf 34},
2423--2434 (1995);  J. Math. Phys., {\bf 36},
No.12, 6625--6632 (1995);
A.Yu. Khrennikov, {\it $p$-adic valued distributions in 
mathematical physics.} Kluwer Academic Publishers, Dordrecht (1994);
A.Yu. Khrennikov, {\it Non-Archimedean analysis: quantum
paradoxes, dynamical systems and biological models.}
Kluwer Acad.Publ., Dordreht, The Netherlands, 1997;

[8] D. Greenberger, M. Horne, A. Zeilinger, Going beyond Bell's theorem,
in {\it Bell's theorem, quantum theory, and conceptions of the universe.} Ed. M.Kafatos,
Kluwer Academic, Dordrecht, 73-76 (1989).

[9] A. N. Kolmogorov, S.V. Fomin, {\it Introductory real analysis.} (Dover Publ. INC, New York) 1975.

[10] A. N. Shiryayev, {\it Probability.} (Springer, New York-Berlin-Heidelberg, 1991).

[11] J. Bell, Introduction to the hidden variable questions. In "{\it Foundations of 
Quantum Mechanics."} (ed. B. D'Espagnat), Course 49, p.171 (Academic Press, New-York, 1972).

[12] Yu. L. Daletskii, S.V. Fomin, {\it Measures and differential equations on 
infinite-dimensional spaces.} (Mir, Moscow, 1983).

[13] D. Bohm  and B. Hiley, {\it The undivided universe:
an ontological interpretation of quantum mechanics.} (Routledge and Kegan Paul, 
London) 1993.

\end{document}